# Photonic realization of a generic type of graphene edge states exhibiting topological flat band

Shiqi Xia[1,*], Yongsheng Liang[1,*], Liqin Tang[1,2], Daohong Song[1,2+] Jingjun Xu[1], Zhigang Chen[1,2+]

*1 The MOE Key Laboratory of Weak-Light Nonlinear Photonics, TEDA Institute of Applied Physics and School of Physics,*

*Nankai University, Tianjin 300457, China*

*2 Collaborative Innovation Center of Extreme Optics, Shanxi University, Taiyuan, Shanxi 030006, China*

*\*These authors contributed equally to this work.*
*+ songdaohong@nankai.edu.cn, zgchen@nankai.edu.cn*

**Abstract:**

Cutting a honeycomb lattice (HCL) can end up with three types of edges (zigzag, bearded and armchair), as is well known in the study of graphene edge states. Here we theoretically investigate and experimentally demonstrate a class of graphene edges, namely, the twig-shaped edges, using a photonic platform, thereby observing edge states distinctive from those observed before. Our main findings are: (i) the twig edge is a generic type of HCL edges complementary to the armchair edge, formed by choosing the right primitive cell rather than simple lattice cutting or Klein edge modification; (ii) the twig edge states form a complete flat band across the Brillouin zone with zero-energy degeneracy, characterized by nontrivial topological winding of the lattice Hamiltonian; (iii) the twig edge states can be elongated or compactly localized along the boundary, manifesting both flat band and topological features. Such new edge states are realized in a laser-written photonic graphene and well corroborated by numerical simulations. Our results may broaden the understanding of graphene edge states, bringing about new possibilities for wave localization in artificial Dirac-like materials.

Graphene, a carbon-based monolayer material, has attracted immense attention due to its fundamental interest and highly exploited applications [1]. Edge-dependent electronic states, for example, have been extensively studied, from the physical properties of graphene nanoribbons [1-4] to possible development of graphene-based spintronic devices [5, 6]. Apart from two-dimensional graphene materials, various synthetic honeycomb lattices (HCLs) have been employed as artificial graphene for electrons, atoms and photons [7], which emulates the behavior of electrons in graphene but avoid the limitation and structure instability of real materials. In particular, photonic graphene, an HCL of evanescently coupled waveguide arrays [8], has been proposed and demonstrated as an ideal platform for investigation of graphene edge states [9, 10], as also implemented in a variety of other settings such as coupled polariton micropillars [11, 12] and microwave resonators [13, 14]. Perhaps, the most exemplary success of photonic graphene-like structures lies in their applications for the exploration of topological physics [15-19], from demonstration of photonic Floquet topological insulators [15] and valley Hall topological insulators [20, 21], to realization of topological surface-emitting lasers [22-24].

Thus far, three types of edges in graphene have been studied: the zigzag, bearded and armchair edges [4]. It is well-known that, while the defect-free armchair edge has no edge state, the zigzag and bearded edges have a set of nearly degenerate edge states, and their eigenvalue spectra are complementary to each other in the one-dimensional Brillouin zone (BZ) [1, 4]. Such electronic edge states of graphene have been observed, but mostly at the zigzag edge or the structured armchair edges [25-29]. Although the bearded (Klein) edge states have never been observed in real graphene owing to the mechanical instability of the dangling carbon, they were realized in the optical analog of graphene – the photonic graphene [10]. Theoretical analyses have shown that these edge states are topologically characterized by the winding properties of the bulk Hamiltonian [30, 31]. In addition, by edge modification of graphene nanoribbons, other edge geometries including decorated armchair edges with Klein nodes have also been proposed [32, 33], but the topological properties of associated edge states and their relations with three well-known edges have hardly been explored. Therefore, a natural question arises: is there a fourth type of edge distinct from the zigzag, bearded and armchair edges? If is, will it support edge states with eigenvalue spectra complementary to the armchair edge?

In this work, in contradistinction to the widely investigated edges in graphene, we propose and demonstrate a new type of edge, which we shall name it "twig" edge due to its shape. Such an edge appears to be a bearded armchair, but we argue that it is a generic type of edge in HCLs by considering the primitive-cell generator rather than a simple lattice truncation of edge modification. We show that the twig edge enables a set of zero-energy edge states over the entire one-dimensional BZ except at the gap closing point ($k = 0$), complementary to the armchair edge in both real-space edge structure and momentum-space eigenvalue spectrum. Moreover, such twig edge states form a one-dimensional dispersionless flat band, characterized by the topologically nontrivial winding of the bulk Hamiltonian. Experimentally, we establish a photonic graphene with twig edges using the laser-writing technique and realize, to the best of our knowledge, the first observation of both elongated and highly compact edge states at the twig edge without employing any defect or nonlinearity.

We first discuss the formation of different edges of graphene (Fig. 1(a)) along with their edge state properties. One kind of unit-cell for the HCL structure is sketched in the lower left corner of Fig. 1(b) (red rhombus) with two assigned basic vectors $(\boldsymbol{a_1}, \boldsymbol{a_2})$. For an infinite lattice (or under a periodic boundary condition), the bulk Bloch Hamiltonian for this chosen unit-cell is

$$\mathcal{H}_{BA}(\boldsymbol{k}) = \begin{bmatrix} 0 & \Delta_{BA}(\boldsymbol{k}) \\ \Delta^*_{BA}(\boldsymbol{k}) & 0 \end{bmatrix}, \quad \Delta_{BA}(\boldsymbol{k}) = t_1 + t_2 e^{ika_2} + t_3 e^{ika_1} \qquad (1)$$

where $t_1$ is the intra-cell coupling and $t_2$, $t_3$ are the inter-cell couplings, the subscript BA indicates *bearded* edge and *armchair* edge. By translating the unit-cell, the HCL with these two edges can be obtained (see the left and bottom edges in Fig. 1). The band structure along the $k_x$ (or $k_y$) direction is calculated by properly choosing the vertical (or horizontal) supercell marked by a shaded rectangle in Fig. 1a. As has been shown before [4], the bearded edge supports edge states with an eigenvalue spectrum residing between two Dirac points in the one-dimensional BZ, whereas the armchair edge has no edge state. On the other hand, one can also have a different choice of unit-cell as illustrated in the upper right corner of Fig. 1(b) (yellow rhombus), where the directions of the basic vectors are changed to $(\boldsymbol{a_3}, \boldsymbol{a_4})$. Translating this unit-cell leads to the formation of HCL with zigzag and twig edges (see the right and top edges in Fig. 1). Similar to Eq. 1, the associated bulk Bloch Hamiltonian $\mathcal{H}_{ZT}(\boldsymbol{k})$ for the *zigzag* and *twig* edges now becomes:

$$\mathcal{H}_{ZT}(\boldsymbol{k}) = \begin{bmatrix} 0 & \Delta_{ZT}(\boldsymbol{k}) \\ \Delta^*_{ZT}(\boldsymbol{k}) & 0 \end{bmatrix}, \quad \Delta_{ZT}(\boldsymbol{k}) = t_1 e^{ika_3} + t_2 + t_3 e^{ika_4} \qquad (2)$$

where $t_2$ becomes the intra-cell coupling and $t_1$, $t_3$ are the inter-cell couplings. Note that each unit-cell and the corresponding Bloch Hamiltonian describe two distinct edge conditions, which are revealed by the supercell selection along different directions.

Apart from the three conventional types of edges, in this work we focus our analysis on the twig edge and its relations with others. The properties of twig edge are described by $\mathcal{H}_{ZT}$. The associated lattice structure is illustrated in the top-right part in Fig. 1 with zigzag edge along the $y$-direction and twig edge along the $x$-direction. The zigzag edge states are denoted by the red lines in momentum space (Fig. 2(a1)), with a spectrum region complementary to that of the bearded edge states [4, 10]. On the other hand, as shown in Fig. 2(a2), the twig edge supports a unique topological flat band constructed by the edge states across the entire one-dimensional BZ. The energy of twig edge states is only distributed at the $B$ sublattices on the top boundary and exponentially decays into the bulk (Fig. 2c). Such zero-energy edge states in graphene are attributed to the bulk topological properties and can be analytically derived from the 'bulk-boundary correspondence' [31, 34]. For the HCL with chiral symmetry, the topological invariant can be described by the winding number [30]:

$$w = \frac{1}{2\pi} \oint \frac{d}{d\boldsymbol{k}} \arg(\Delta(\boldsymbol{k})) d\boldsymbol{k} \qquad (3)$$

where $\boldsymbol{k}$ can be $k_x$ or $k_y$ component depending on the edge direction, and $\Delta(\boldsymbol{k})$ is the off-diagonal term of bulk Hamiltonian. Considering the zigzag and twig edges, the bulk Hamiltonian can be rewritten as $\mathcal{H}_{ZT}(\boldsymbol{k}) = \boldsymbol{h}_{ZT}(\boldsymbol{k}) \cdot \boldsymbol{\sigma}$, where $\boldsymbol{h}_{ZT}(\boldsymbol{k}) = (\text{Re}(\Delta_{ZT}(\boldsymbol{k})), \text{Im}(\Delta_{ZT}(\boldsymbol{k})))$, and the Pauli matrix $\boldsymbol{\sigma} = (\sigma_x, \sigma_y)$. To calculate the winding number for the zigzag edge, the Bloch vector $k_y$ is fixed and the orientation of $\boldsymbol{h}_{ZT}(\boldsymbol{k})$ varies along $k_x$, as illustrated by the blue arrows in Fig. 2(b). If $\boldsymbol{h}_{ZT}(\boldsymbol{k})$ rotates to make a whole loop within a period of $k_x$, $w$ is nonzero which leads to the presence of edge states at the given $k_y$. The shaded regions in Fig. 2(b) demonstrate the values of $k_y$ for which nontrivial winding leads to the zigzag edge states, consistent with the results in Fig. 2(a1). To directly illustrate their winding path, $\boldsymbol{h}_{ZT}(\boldsymbol{k})$ with different $k_y$ is plotted in the $(\sigma_x, \sigma_y)$ plane in Fig. 2(d). The origin point $\mathcal{O}$ marked by a red dot in Fig. 2(d2, e1) is the gap closing and phase transition point [35]. When $\mathcal{O}$ is encircled by the winding loop (Fig. 2(d3)), it falls in the shaded region in Fig. 2(b) where the edge states exist with a nonzero winding number. The other two loops in Fig. 2(d1, d2) correspond to the trivial and transitional cases at $k_y = 0$ and $k_y = 2\pi/3a$. On the other hand, when $\boldsymbol{h}_{ZT}(\boldsymbol{k})$ is employed to analyze the topological properties for the twig edge, $k_y$ varies constantly for any $k_x$, and an example at $k_x = \pi/\sqrt{3}\,a$ is shown by the vertical line in Fig. 2(b). The winding loops always encircle $\mathcal{O}$ in the $(\sigma_x, \sigma_y)$ plane, except at the gap closing point at $k_x = 0$ (Fig. 2(e)). As such, the twig edge states exist and are characterized by nontrivial winding throughout the whole one-dimensional BZ, forming a topological flat band.

We emphasize our approach for understanding the relation between bulk topological properties and edge conditions in graphene: to unveil the topological nature of edge states, the edges should be considered from "piling up" the unit-cells instead of specific truncations of the HCL. Thus, the boundary sites must be contained within a complete unit-cell, which determines the basic vectors and the corresponding bulk Hamiltonian. For instance, by comparing $\mathcal{H}_{BA}$ and $\mathcal{H}_{ZT}$, one can see $\Delta_{ZT}(\mathbf{k}) = \exp(i\mathbf{k}\mathbf{a}_3) \Delta_{BA}(\mathbf{k})$ due to the different orientations of basic vectors. When only the bulk property of the system is considered, both Hamiltonians can be employed to describe the properties of the bulk and the gauge term $\exp(i\mathbf{k}\mathbf{a}_3)$ has no physical effect. However, once the edges are considered, one unique bulk Hamiltonian is written to satisfy the edge condition. Moreover, the topological features along $k_x$ and $k_y$ are affected by the gauge term (Eq. 3), which in turn determines the existence of edge states. Note that the gauge dependence of edge states does not conflict with the gauge invariant of the Berry phase which is derived from a given bulk Hamiltonian. Hence, the interplay between the edge conditions and the winding properties of bulk Hamiltonian reveals the distinctive feature of zero-energy edge states with chiral symmetry.

Next, we present experimental realization of twig edge states using photonic graphene. The photonic HCL with the desired edge structure is established by the cw-laser writing technique in a nonlinear crystal (SBN) [36]. Such an HCL with two twig (top and bottom) and zigzag (left and right) edges is shown in Fig. 3(a), with a lattice spacing of $32\ \mu m$. To excite the twig edge states, a probe beam with specific amplitude and phase distribution matching that of the theoretically calculated edge mode (Fig. 2(c)) is launched into the HCL at the top edge [35]. To show that the twig edge states exist in a large momentum region, the probe beam with two different transverse momenta is selected, matching the blue- (green-) point excitation at $k_x = \pi/\sqrt{3}a$ ($k_x = \pi/2\sqrt{3}a$) in Fig. 2(a2). It should be noted that the zigzag or the bearded edge cannot support edge states simultaneously at these two relative spectral positions in the one-dimensional BZ. At the twig edge, however, the probe beam remains localized at the edge and populates only the B sublattices after 20mm of propagation through the lattice (Fig. 3(b1, b2)), forming the edge states in agreement with calculated results (Fig. 2(c1, c2)). At the two ends (outside of the dashed square), there is leakage of energy to the A sites, which is mainly due to the finite-size effects along the *x*-direction. For comparison, the mixed bulk modes excitation at $k = 0$ are sent to the edge [35]. The output of the probe beam with mixed bulk modes cannot be localized on the edge and couples into the *A* sublattice (Fig. 3(b3)), which is in contrast to that in Fig. 3(b1-b2). To explicitly show the excitation condition, the Fourier spectra of the probe beam are experimentally captured and shown in Fig. 3(d) for the blue- and green-point excitations, corresponding to the edge and middle spectral components of the one-dimensional BZ, respectively. Since the propagation distance in experiment is limited by the crystal length, numerical simulations for a longer propagation distance (50mm) are presented in Fig. 3(c) to better differentiate the dynamics: for the edge mode excitation, light remains edge-localized, but for the mixed mode excitation it spreads into the bulk occupying both A and B sublattice sites. More experimental results can be found in Supplementary Materials [35].

Since all twig edge states are degenerate in the whole one-dimensional BZ and form a dispersionless flat band, it enables the formation of compact edge state along the boundary by appropriate superposition of edge states with different momenta [37-40]. We demonstrate the edge excitation at the zigzag and twig edges for comparison, which share the same bulk Hamiltonian but the zigzag edge states occupy only a portion of the one-dimensional BZ. To show the compactness of the edge state in different edge conditions, it is convenient to select a site along the edge and remove all the mode components of dispersive bands. Figure 4 (a1, c1) is the distribution of the modified edge states for twig and zigzag edges. Most of light along twig edge is located in one supercell (Fig. 4(a2)), while that of the zigzag edge state elongates to neighbor sites along the edge (Fig. 4(c2)). The mode mapping $R = |\langle\varphi_k|\Psi\rangle|^2$ is defined to show the distribution of such edge state in momentum space, where $\Psi$ is the modified edge state shown in Fig. 4(a1, c1) and $\varphi_k$ denotes the edge eigenstate at momentum $k$. The modified edge state along the twig edge spans the entire one-dimensional BZ (Fig. 4(a3)), indicating a more compact form along the boundary in real space compared to that at the zigzag edge (Fig. 4(c3)). We call the edge state shown in Fig. 4(a1) "compact edge state". To demonstrate such a distinctive compact feature of twig edge states in experiment, the probe beam which matches the compact edge state (Fig. 4(b1)) is sent to the lattice, and it remains intact and compact after 20mm propagation (Fig. 4(b2)). On the other hand, when the same probe beam excites only one supercell along the zigzag edge (Fig. 4(d1)), since zigzag edge cannot support the compact edge state, light spreads into *A* sublattices (Fig. 4(d2)). These experimental results illustrate the unique feature of the compact edge state originating from the topological flat band formed by the degenerated edge eigenstates.

In summary, we have proposed and experimentally demonstrated a new type of edge states using photonic graphene. The new twig edge forms a complete set of edge conditions together with three other well-known edges. The topological origin of the edge states is discussed, unraveling the relation between edge conditions and bulk topological properties. Moreover, soliton-like compact edge states exhibiting both flat band and topological features are observed, with edge states confined in one supercell along the boundary without employing any defect or nonlinearity. Realization of the twig edge and its characteristic edge states in HCLs opens up an avenue for fundamental research in many areas, such as the valley Hall effect, energy confinement, topological corner and edge states [15, 21, 40-44] in Dirac-like systems, which may lead to unconventional applications.

**Acknowledgements**

This work was supported by National Key R&D Program of China (2022YFA1404800), the National Natural Science Foundation of China (12134006, 12274242, 11922408, 12204252), China Postdoctoral Science Foundation (BX2021134, 2021M701790), and the Natural Science Foundation of Tianjin for Distinguished Young Scholars (Grant No. 21JCJQJC00050), PCSIRT (IRT_13R29), 111 Project (No. B07013) in China.

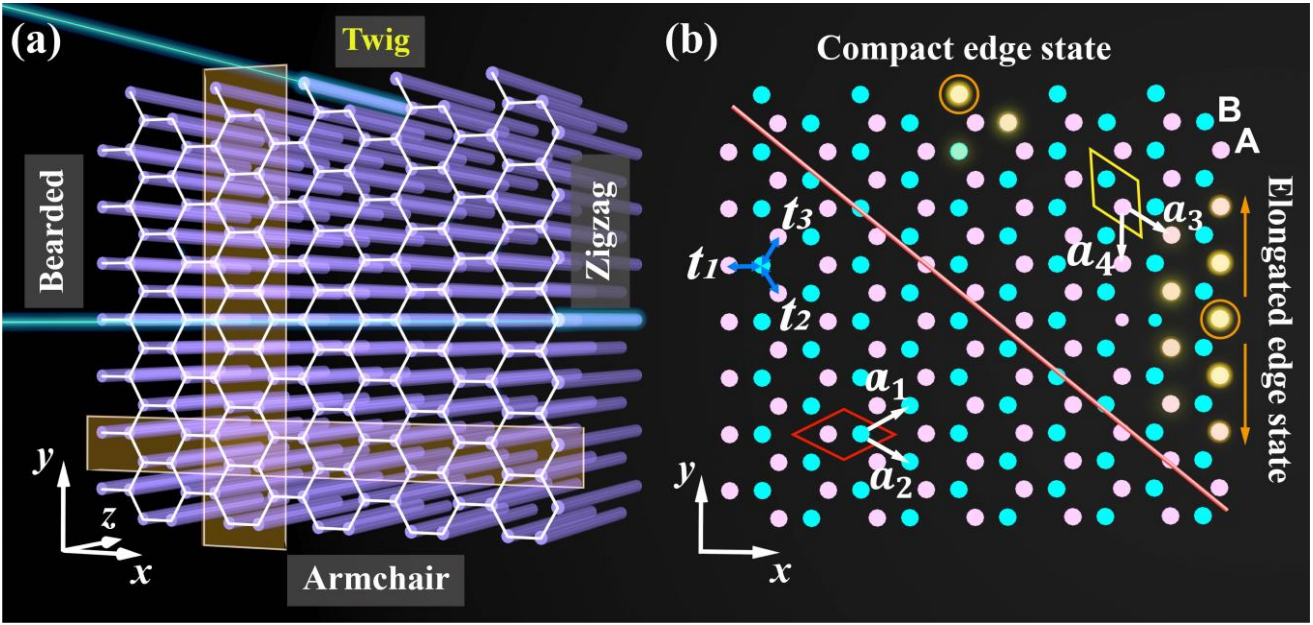

**Fig.1 Illustration of the HCL with four different edge conditions.** (a) Schematic of an HCL (array of waveguides) with four distinct types of edges. Two shaded rectangles illustrate supercells for bearded/zigzag (horizontal) and twig/armchair (vertical) edges. Beams aiming at single site of the twig (zigzag) edge will be compact (elongated) along the edge, indicated by bright yellow dots in (b). (b) HCL structure with two sublattices (A, B) in $(x, y)$ plane. The unit-cell enclosed by a red rhombus is for bearded and armchair edges, and that for zigzag and twig edges is indicated by a yellow rhombus. $a_1, a_2$ and $a_3, a_4$ are the corresponding basis vectors when different unit-cells are selected. $a_1 = \sqrt{3}a/2\,\hat{x} + a/2\hat{y}$, $a_2 = \sqrt{3}a/2\,\hat{x} - a/2\,\hat{y}$ and $a_3 = \sqrt{3}a/2\,\hat{x} - a/2\,\hat{y}$, $a_4 = -a\hat{y}$, $a$ is the lattice constant.

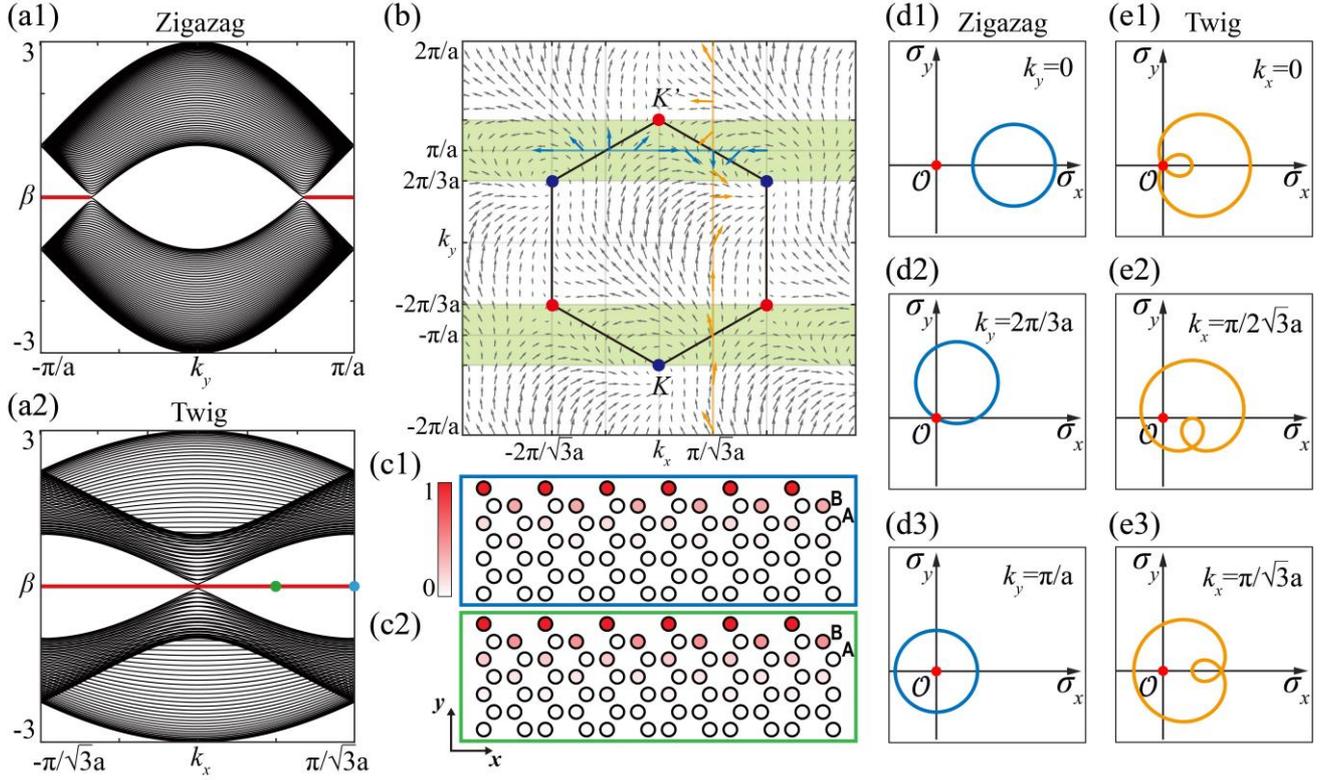

**Fig. 2 Theoretical analysis of the topological property of zigzag/twig edge states in HCLs.** (a) One-dimensional band structures for HCL with (a1) zigzag and (a2) twig edges, where the red lines represent the regions of edge states. (b) Vector fields of $h_{ZT}(k)$. Red and blue dots represent two inequivalent Dirac points $K$ and $K'$ at corners of the first BZ. The values of $k_y$ at the shaded regions indicate the zigzag edge with nontrivial winding, where blue arrows indicate the direction of the vector field along $k_y = \pi/a$. Twig edge states exist for any $k_x$, and orange arrows are for that of the twig edge at $k_x = \pi/\sqrt{3}\,a$. (c) Normalized intensity distributions of twig edge states at (c1) $k_x = \pi/\sqrt{3}a$ and (c2) $k_x = \pi/2\sqrt{3}a$ denoted by blue and green dots in (a2), respectively. (d) Winding loops for the zigzag edge in the $(\sigma_x, \sigma_y)$ plane at different $k_y$. The red dot marks the origin point $\mathcal{O}$. (e) has the same layout as (d) but for the twig edge at different $k_x$.

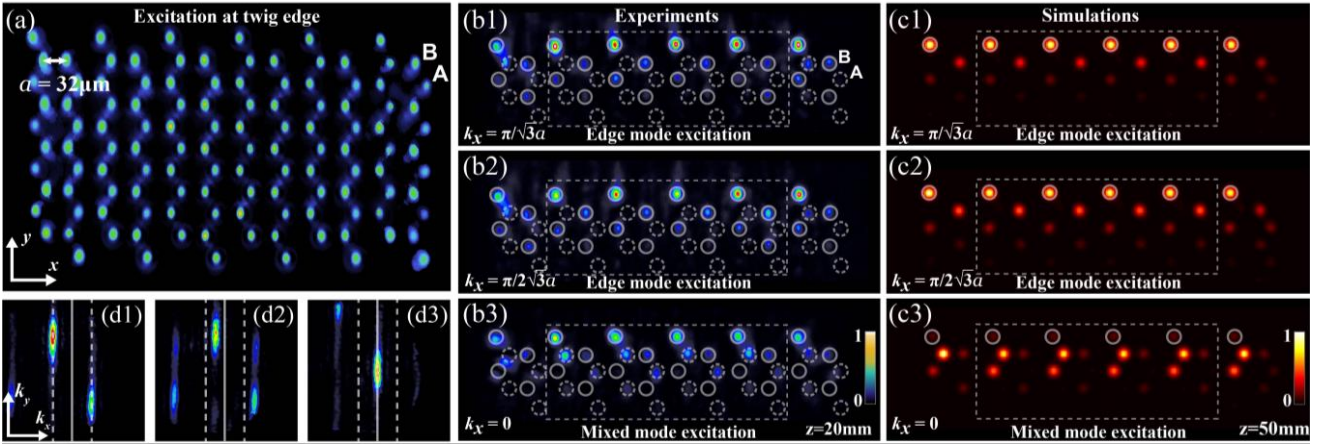

**Fig. 3 Experimental and numerical demonstration of the twig edge states in photonic graphene.** (a) Photonic graphene with twig edges along the $x$-direction in the experiment. (b1, b2) Experimental outputs of the probe beams which match the eigenmode of twig edge states at (b1) $k_x = \pi/\sqrt{3}a$ and (b2) $k_x = \pi/2\sqrt{3}a$. (b3) The output for the mixed mode excitation at $k_x = 0$. The corresponding simulation results for a longer propagation distance ($z = 50mm$) are shown in (c), where the solid white circles mark the sites on the topmost twig edge. (d) Fourier spectra of the input beams corresponding to (b), in which solid (dashed) lines mark the center (edge) of the one-dimensional BZ.

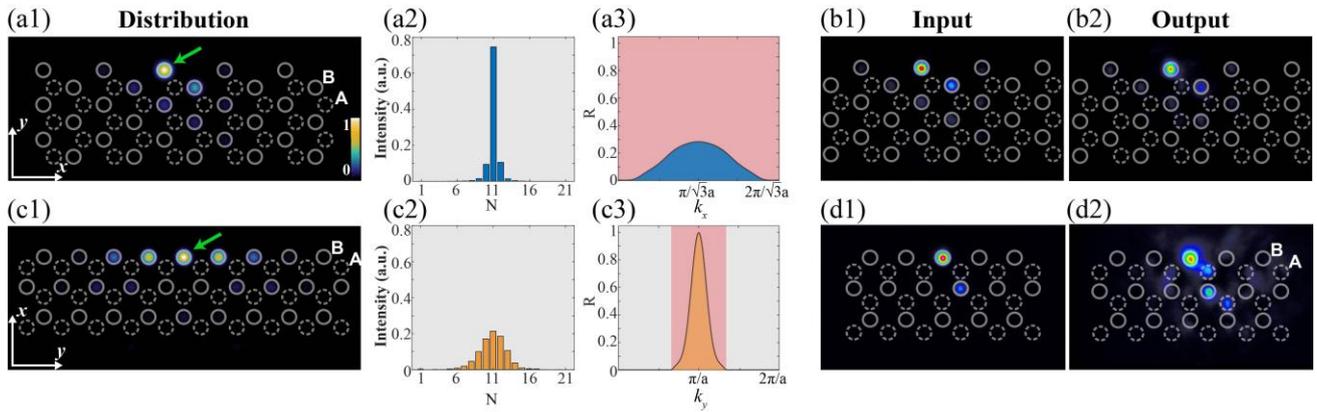

**Fig.4 Demonstration of compact edge states at the twig edge of photonic graphene.** (a) Demonstration of the compact edge state for the twig edge. (a1) The intensity distribution of the modified edge state. The initial selected site is marked by a green arrow. (a2) The percentage of energy projected on supercells ($N$) along the boundary. (a3) The spectrum distribution of the compact edge state in $k$ space, where edge states exist in the red shaded region. (b) Experimental results of compact edge states. (b1) The probe beam matching the compact edge state. (b2) The output of the probe beam after $20\ mm$ propagation. (c-d) have the same layout as (a-b) but for zigzag edge for comparison.